# Coherence spectroscopy by the Nth power of the measured signal in an interferometer overcoming the diffraction limit

Byoung S. Ham[1,2]
[1]School of Electrical Engineering and Computer Science, Gwangju Institute of Science and Technology, 123 Chumdangwagi-ro, Buk-gu, Gwangju 61005, South Korea
[2]Qu-Lidar, 123 Chumdangwagi-ro, Buk-gu, Gwangju 61005, South Korea
(May 20, 2024; bham@gist.ac.kr)

**Abstract**
Coherence spectroscopy has been intensively studied over the last several decades for various applications in science and engineering. The Rayleigh criterion defines the resolution limit of an interferometer, where many-wave interference beats the resolution limit of a two-slit system. On the other hand, the diffraction angle through a slit is reduced by the Kth power of the measured signal, resulting in the shot-noise limit. Here, the Kth power of the measured signal in an N-slit interferometer is studied for enhanced coherence spectroscopy to overcome the resolution limit of the original system. The Kth power to the individual intensities of the N-slit interferometer is numerically demonstrated for enhanced resolution satisfying the shot-noise limit. As a result, the Kth power of the intensity beats the resolution limit of the N-slit interferometer, in which the out-of-shelf spectrometer or wavelength meter can be a primary beneficiary of this technique. Due to the same resolution ($\pi/N$) of the Heisenberg limit in quantum sensing as in the N-slit interference fringe, the proposed Kth power technique also beats the superresolution in quantum metrology.

**Introduction**
High-resolution spectroscopy has been intensively studied for various applications in science and technologies over the last several decades [1-13]. The diffraction limit or Rayleigh criterion classically determines the resolution limit of an optical signal measured in an interferometer. Thanks to the quantum superposition of multiple waves, an N-slit or N-groove grating-based interferometer achieves enhanced resolution, whose resolution limit reaches $\pi/N$ in phase space [14]. Although the N-slit resolution enhancement seems to satisfy the Heisenberg limit in quantum sensing [11-13], the physics is completely different due to the first-order limited intensity correlation [14,15]. Recently, a technique of higher-order intensity product via projection measurements has been proposed [16] and experimentally [17] demonstrated for the shot noise limit (SNL) in a Mach-Zehnder interferometer (MZI). Due to the statistical ensemble of measured signals, the demonstrated intensity correlation of the MZI output field satisfies the definition of SNL. Due to the incoherence nature of the intensity correlations demonstrated for SNL [17], the Kth power to the fringe of an N-slit interferometer should work for the same resolution enhancement beyond the Heisenberg limit ($\pi/N$) in quantum sensing.

The higher-order intensity product of the interference fringes via projection measurement affects the resolution limit defined by the Rayleigh criterion [16,17]. Whether the nature of the ordered intensity products is coherent or not, the coherence spectroscopy of an interferometer can result in either superresolution [18-20] of quantum sensing or classical SNL [16,17]. In both cases, the diffraction angle of light is reduced resulting in resolution enhancement [15]. Here, the same physics of SNL that was recently demonstrated for MZI fringes via projection measurements [17] is now applied to the N-slit interferometer to show enhanced resolution beyond the diffraction limit given by the Rayleigh criterion [14]. For this, first, the Kth power of the MZI output signal is numerically demonstrated for the same SNL as in the projection measurements of the recently demonstrated [16,17]. Then, the Kth power to the N-slit interferometer is numerically calculated for enhanced resolution beyond the Rayleigh [14] and the Heisenberg limit [11-13,20]. In addition, the Kth power resulting in enhanced spectroscopy intrigues a philosophical issue because the post-measurement never intervenes in the interference process (see Discussion).

**Results**

Figure 1 shows the schematic of projection measurements for a scalable Kth-order intensity correlation satisfying SNL [16]. Recently, the same intensity product has been experimentally demonstrated for up to the fourth-order of the MZI output signal, whose resolution enhancement is the square root of the intensity-product order K. [17]. Due to the Born's rule stating that intensity $(I)$ is the absolute square of the amplitude $(E)$, however, all divided output fields used for the projection measurement in Fig. 1 are treated as $I_2/K$, where K is the number of divided output ports by the balanced nonpolarizing beam splitter (BS). Thus, the statistical ensemble is satisfied for independent and incoherent photon nature interacting in the interferometer. According to quantum mechanics, the number of split ports is up to the photon number of the input light [11,18-20,21]. The projection measurement technique has already been adapted for quantum sensing using maximally entangled photon pairs of N00N state [21,22]. Due to the statistical events, thus, the projection measurement in Fig. 1 can be replaced by the Kth power $I_2^K$ of the MZI output intensity $I_2$. As a result, the complicated optical system by BSs in Fig. 1 can be simplified by a single logic gate.

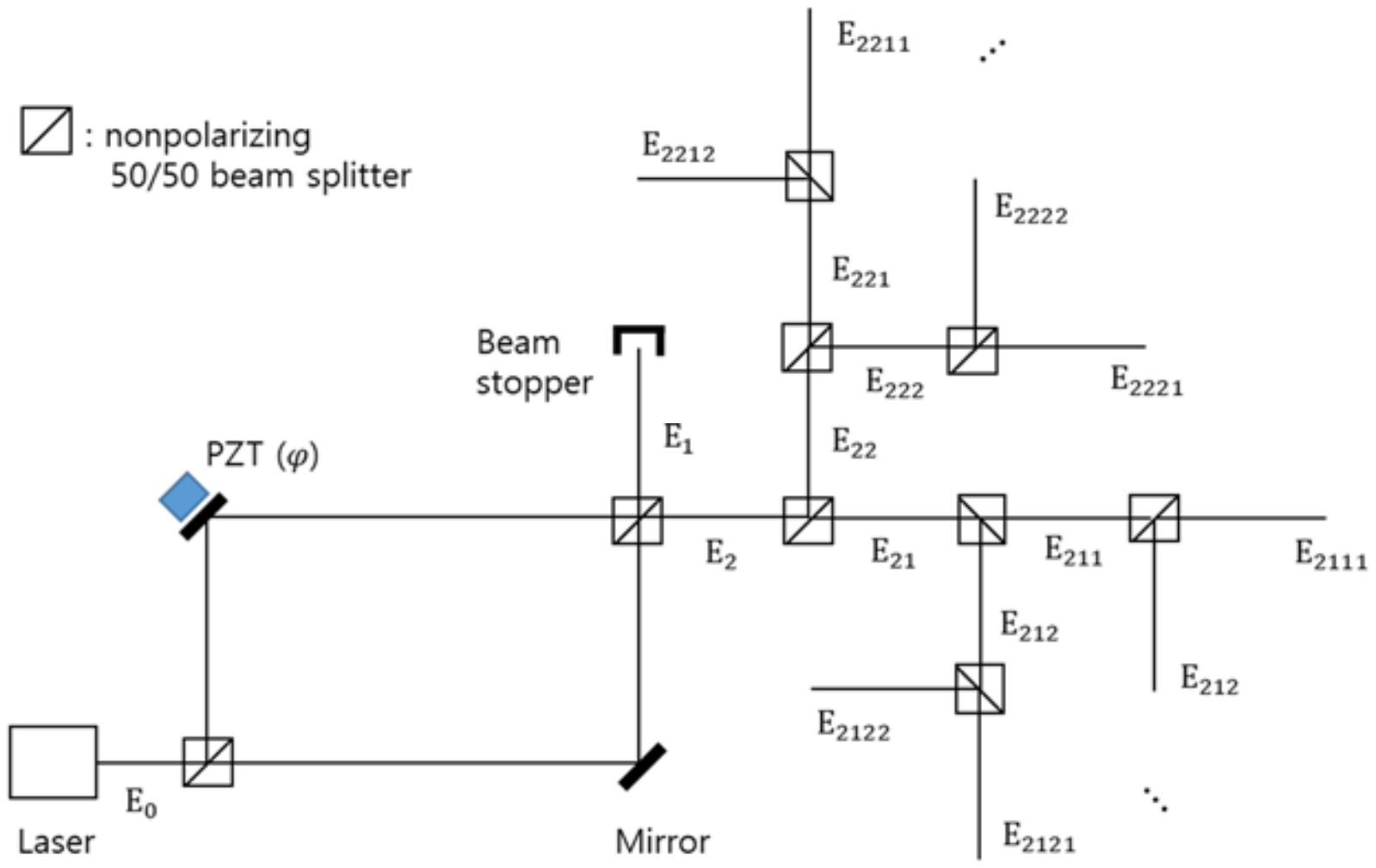

**Figure 1.** Schematic of projection measurements, satisfying shot noise limit.

Figure 2 shows numerical simulations of the proposed Kth power to the output signal from an N-slit ($2 \leq N \leq 200$) interferometer, where the N-slit fringes satisfy the classical resolution $\pi/N$ by the Rayleigh criterion [14]. The left panel of Fig. 2 is for the numerical simulations of the first-order intensity correlation $I_N$ of the N-slit interferometer as a function of the phase variation $\varphi$ and the slit number N. The right panel confirms the N-slit resulting resolution, satisfying $\pi/N$, by definition of the Rayleigh criterion [14].

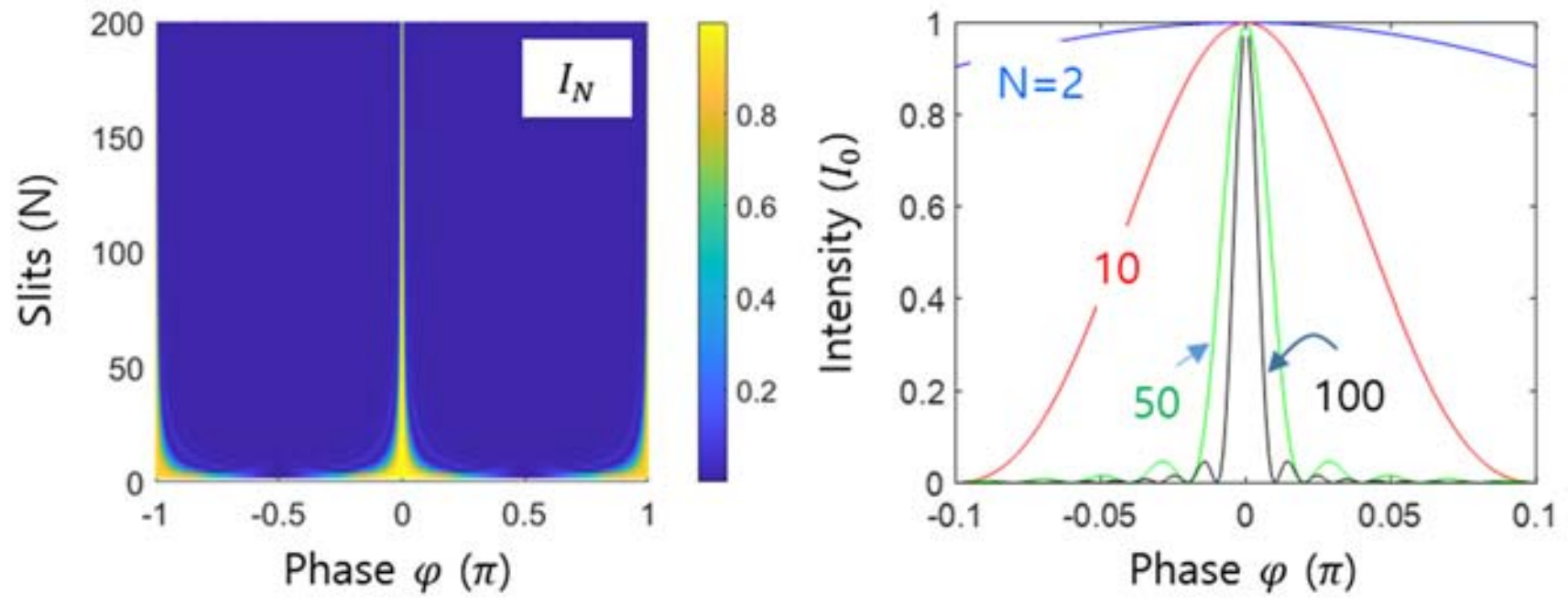

**Figure 2.** Numerical calculations for N-slit interference. $\beta = 6\alpha$ (See the text). (left panel) Output intensity of an N-slit interferometer. (right panel) N-dependent resolution. The number indicates slits.

As a general solution, the discretely increased phases among N waves resulting from the N slits contribute to the resolution enhancement via many-wave interference $I_N = (sinc\beta)^2 \left(\frac{sinN\alpha}{sin\alpha}\right)^2$, where $\beta = \mathrm{kbsin}\theta/2$ and $\alpha = \mathrm{kasin}\theta/2$ [14]. In $I_N$, 'a' and 'b' represent the slit-to-slit distance and slit width, respectively. The first term $sinc^2\beta$ represents the single slit's diffraction effect, while the second term is for the N-slit interference contributing to the enhanced resolution (see the right panel in Fig. 2). This type of resolution enhancement is linearly proportional to the number of fields (or slits) and thus is the same as superresolution in quantum sensing for the Nth-order intensity correlation in MZI [18-22]. In coherence spectroscopy, the intensity correlation in Fig. 1 satisfies SNL, though (see Fig. 3) [16,17].

Figure 3 shows numerical calculations of the Kth-order intensity product for the K-divided output fields of an N-slit interferometer, as in Fig. 1. In Fig. 3, the Kth-order intensity correlation is analyzed for the N-slit interferometer not only for the enhanced resolution to beat the diffraction limit in classical physics but also for the discussion of post-measurements of the interference fringes. This type of intensity product can be performed to the optical-to-electric (OE)-converted signal due to the same (indistinguishable) intensities. For this, three different slit numbers are considered for N=2, 10, and 100. The resulting Kth power $I_{N=2}^{k}$ in Fig. 3(a) shows K-dependent line narrowing, where $1 \leq \mathrm{K} \leq 100$. Figure 3(b) is for the same K applied to a ten-slit (N=10) interferometer, where the Inset is for N=100.

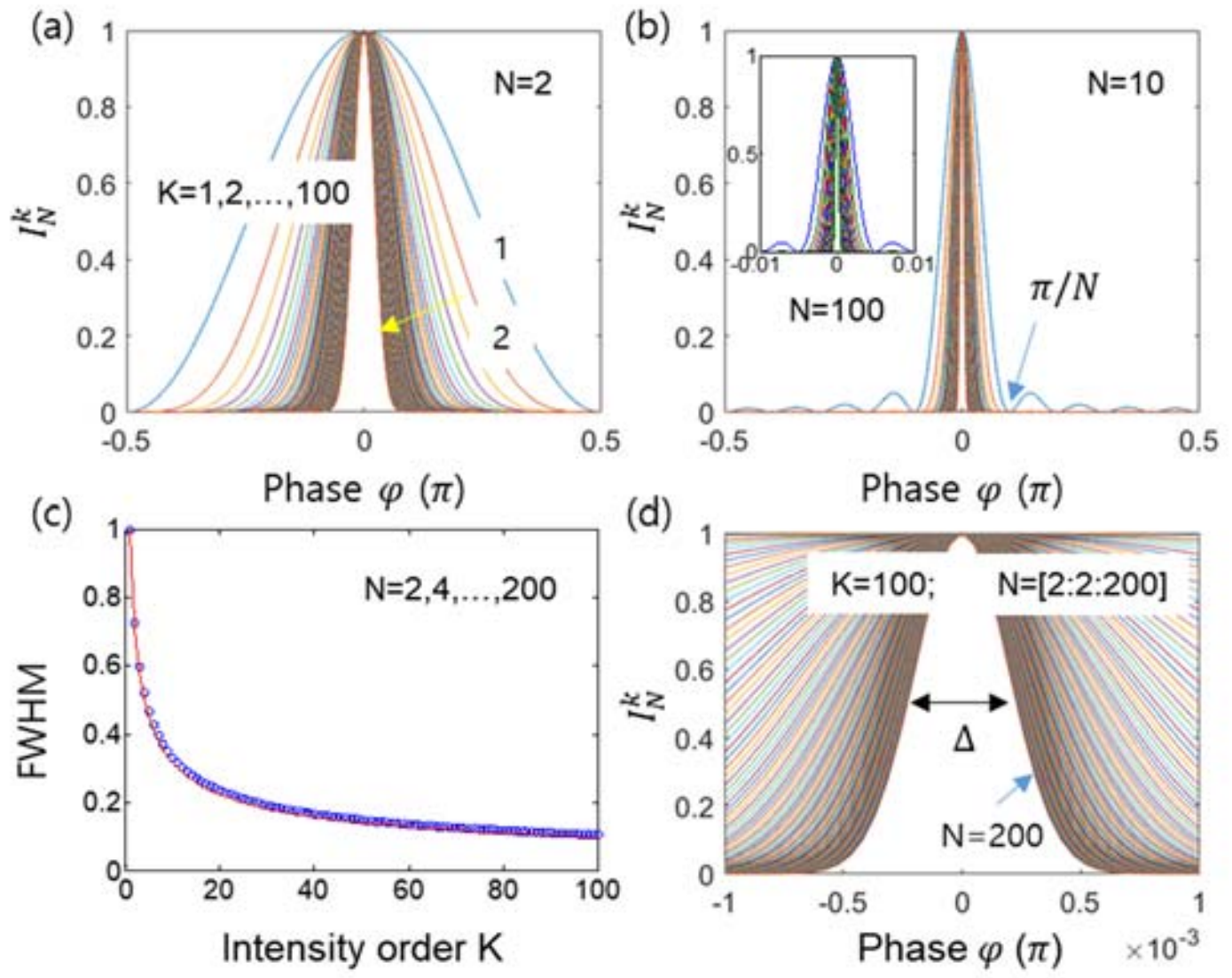

**Figure 3.** Numerical calculations of the intensity products for N-slit interferometer in Fig. 1. (a) For a two-slit interferometer. (b) For N=10. Inset: N=100. (c) Calculated resolution (FWHM) for N=j (2,4,...,200). (d) K=100 for all N. The K is the intensity-product order. N is the number of slits.

Figure 3(c) shows the Kth power resulting resolution enhancement satisfying SNL by $\sqrt{K}$. For this, the full-width at half maxima (FWHM; $\Delta_N$) of all curves in Figs. 3(a) and (b) are individually measured using a homemade program. The open circles are for the measured numerical data, while the red curve is for SNL. The small deviation between them is due to ideal sine function in $I_N$ based on monochromatic waves. For the statistical nature, signal or events must be Gaussian distributed. In real laser interference, of course, the Gaussian distributed photon characteristics are satisfied by the Poisson statistics for high mean photon number. The SNL feature of the Kth power is satisfied for all N in Fig. 3(c), resulting in the $\sqrt{K}$-enhanced resolution for any N-slit interferometer.

Figure 3(d) is for a fixed intensity product with K=100 applied to all Ns (2,4,...,200). By the Kth power, the same resolution enhancement by SNL for all Ns is numerically confirmed (see $\Delta = 10^{-3}\pi/2$ for N=200).

Compared to the resolution of $\pi/100$ for N=100 in the right panel of Fig. 2, the Kth-power $I_{200}^{100}$ (N=200; K=100) shows $2\sqrt{100}$ times enhanced resolution, resulting in $\pi/2000$. Thus, the Kth-power of N-slit interference fringes is well presented for an arbitrary N, satisfying SNL, whose resolution enhancement is the square root of K for all N.

Figure 4 is for a two-slit interferometer or MZI for the frequency variation with $\delta f = \pm 0.2 f_0$. As discussed in Fig. 2 for an N-slit interferometer, the resolution of the two-slit system is fixed at $\pi/2$. Depending on the ratio of 'b' to 'a' in the slit configuration, the diffraction term $sinc^2\beta$ results in decreased $I_N$ as the phase difference $\varphi$ increases. Around $\varphi = 0$, however, $I_N$ has nearly no phase shift, as shown in Fig. 4(a). Due to the phase 'α' in $I_N$, where $\alpha = 2\pi f \delta T$, thus, the fringe of $I_N$ is determined by the product between frequency f and delay time $\Delta T$ (see Fig. 4(a)). For this reason, the minimum resolvable frequency difference can be found numerically. Figure 4(b) shows an example of unresolvable frequencies $f_0$ and $f'(= 0.8 f_0)$ according to the Rayleigh criterion [14].

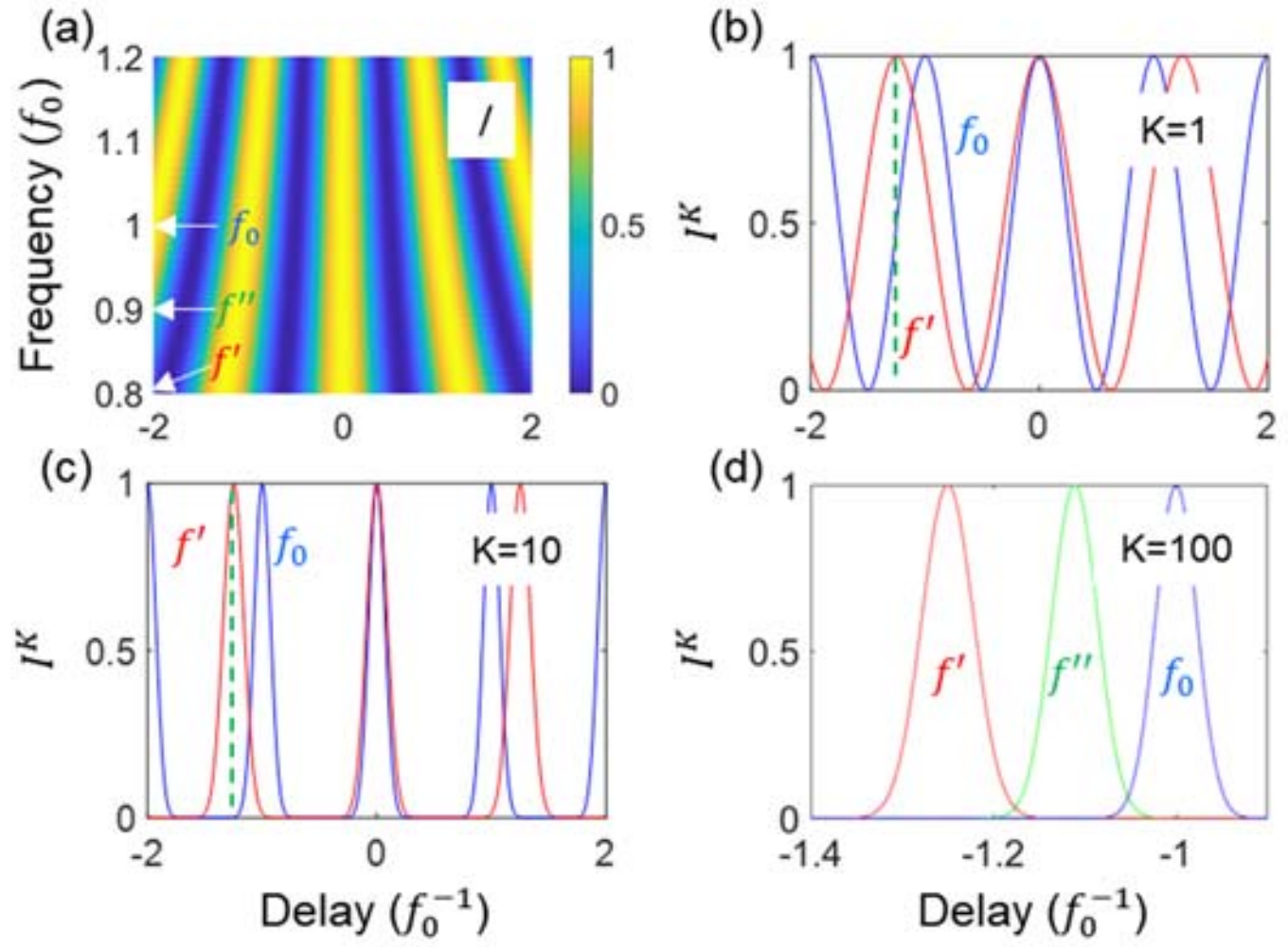

**Figure 4.** Numerical calculations of the frequency-dependent interference fringes in a double-slit system (N=2) or Mach-Zehnder interferometer. (a) and (b) K=1. (c) K=10. (d) K=100.

Figures 4(c) and (d) are for the Kth power applied to Fig. 4(a). As shown in Fig. 4(c), the unresolved frequencies in Fig. 4(b) are now turned out to be resolvable by K=10 (see the green dashed line). With K=100 in Fig. 4(d), the frequency resolution is further enhanced by $\sqrt{10}$ by the SNL effect, as analyzed in Fig. 3(c). Thus, the post-control of the interference fringes by the Kth power can contribute to the resolution enhancement of the coherence spectroscopy based on an N-groove grating or Fabry-Perot interferometer (FPI). The practical benefit of the intensity product is simple due to the logic operation and powerful for the high-end application to the out-of-shelf spectrometer or wavelength meter.

Figures 5(a) and (b) are for N=10 and 100 in Fig. 2, respectively, with no Kth power. The resolution enhancement in Figs. 5(a) and (b) is only due to the multi-wave interference, resulting in $\pi/N$, as discussed in Fig. 2 [14]. Figure 5(c) is for K=1 (red; green; blue), 10 (magenta; green dashed; cyan), and 100 (dotted), where the line narrowing of all different frequencies ($f'$, $f''$, $f_0$) follows SNL: 1; $1/\sqrt{10}$; 1/10. Thus, the resolvable frequencies in Fig. 5(d) for N=K=100 are ten times closer to those in Fig. 5(c), where $f''' = 0.99 f_0$. Compared with the same K in Fig. 4(c) but different N=2, Fig. 5(c) shows 50 times enhanced resolution by N=100. Thus, it is well understood that the resolution of an interferometer can be enhanced further beyond the resolution limit $\pi/N$ by the Kth power for SNL. Surprisingly, this Kth power-induced resolution $\pi/(N\sqrt{K})$ in Fig. 5(c) beats the quantum sensing limited by the Heisenberg limit [11-13,18-20]. Thus, the conventional spectrometer or wavelength meter based on an N-groove grating or FPI can be directly applied for the present Kth power technique for better coherence spectroscopy. Due to the post-control of the Kth power to the N-slit interference fringes, this technique can be directly used for an electrical signal of the interferometer.

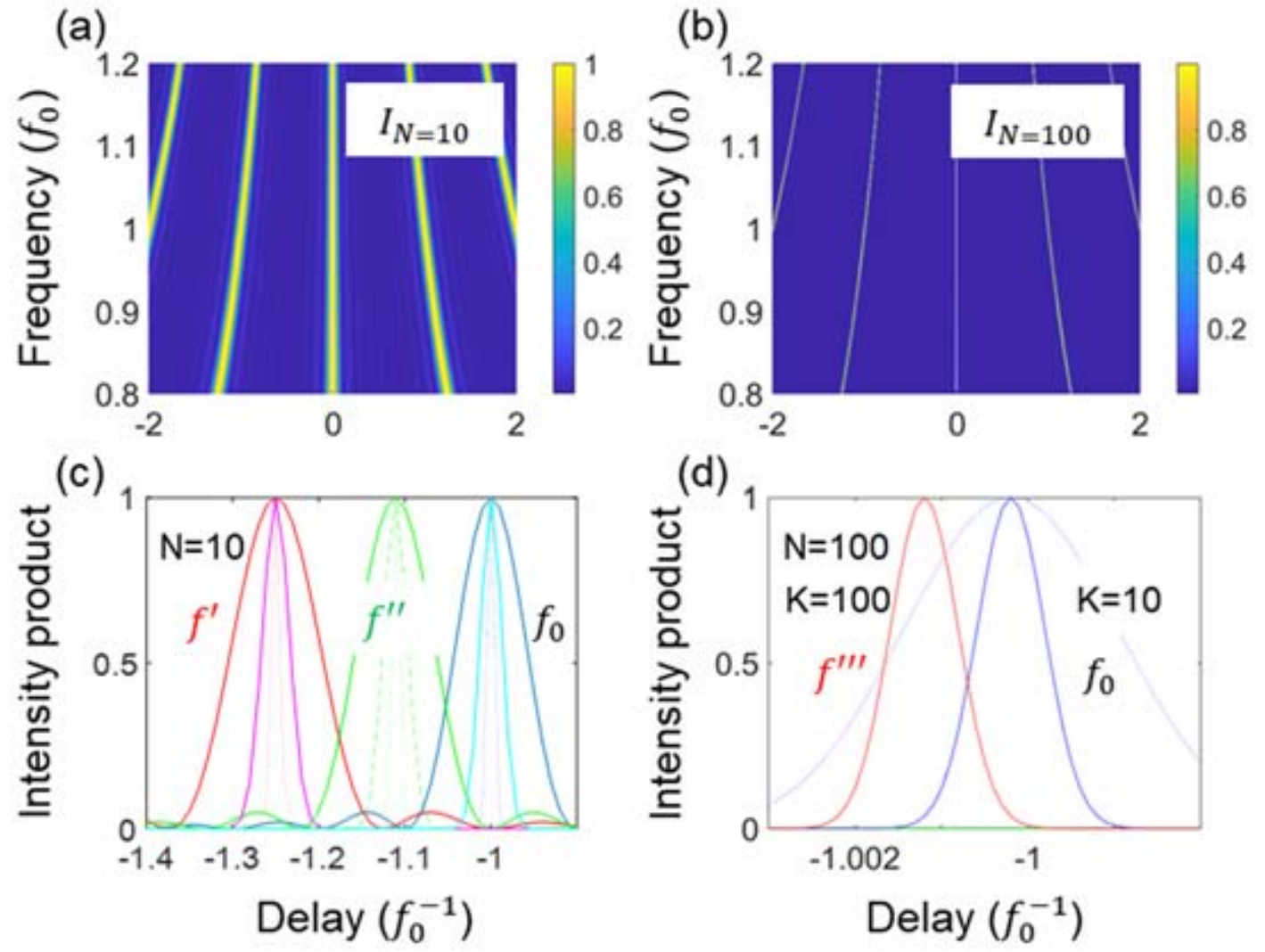

**Figure 5.** Numerical calculations of the N-slit interference. (a) N=10. (b) N=100. (c) K=1;10;100. (d) K=N=100. Dotted: K=10. $f'' = 0.90f_0$. $f''' = 0.99f_0$. $\beta = 6\alpha$ (See the text).

**Discussion**

In conventional N-slit (or N-groove grating) coherence spectroscopy, the proposed intensity-product technique is for a post-measurement of the optical signal from an interferometer. Thus, a quick and important question may be asked how does the intensity product of the electrical signal retrospectively affect the optical interference? In other words, how does the initially unresolved two different frequencies such as in Fig. 3(b) turn out to be resolvable by a simple logic gate operation of the Kth power to the measured signal even without physical intervening in the optical system? Because the N-wave superposition predetermines the resolution of the interferometer, the post-measurement process-dependent enhanced resolution does not sound right. This kind of dilemma has already been raised and discussed for the quantum eraser, even though completely different physics of the wave-particle duality is involved [23-25].

To answer this question, we focus on the classical feature of SNL where redundant measurements of the statistical ensemble reduce the phase error, resulting in resolution enhancement. Here, the Kth power of the interference fringe in Fig. 1 fits the statistical ensemble due to the incoherent nature of the split intensities, where the projection measurement is for the phase bases of the interferometer. Unlike the polarization-basis projection in quantum erasers [21-23] or Bell inequality violation [26] onto a polarizer, the present intensity product between divided optical signal is phase or polarization-independent due to Born's rule, stating that the amplitude's phase has nothing to do with intensity. Due to the Poisson-distributed photons of the laser, Gaussian statistics for SNL is fully satisfied for the measured events in the present macroscopic regime of Fig. 1.

More importantly, the Kth power to the interference fringes results in a 'surprising' breakthrough even in quantum metrology due to the square root enhancement of the N-slit resolution $\pi/N$, where the N-slit resolution is equivalent to the superresolution in quantum sensing [11-13,18-22]. As the quantum sensing is accomplished by the Nth-order intensity correlation between entangled photons [11-13,21,22] or phase-controlled coherent lights [18-20], the discrete phase relation between N waves in the N-slit interferometer should give a similar phase relation between paired photons [19]. Thus, the present Kth power applied to the N-slit interferometer results in the most resolvable spectroscopy beating superresolution in quantum sensing [11-13,18-20]. Using a simple logic gate operation for the Kth power of the electrical signal is a great benefit to the high resolution spectroscopy of an unknown optical field. Due to the SNR, the Kth power to the interference fringe also contributes to noise reduction.

**Conclusion**

Intensity-product-based coherence spectroscopy was presented to beat the diffraction limit of the light in an N-slit interferometer. Numerical simulations of the Kth-order intensity product of the measured signal from an N-slit interferometer were conducted for further enhanced resolution satisfying SNL. Most of all, the Kth power-resulting resolution enhancement was effective for all N-slit-induced interference fringes. Thus, the proposed intensity-product technique can be directly applied to the out-of-shelf spectrometer and wavelength meter by simply adding an electronic chip for the Kth-order intensity product of measured fringes. As discussed in the quantum eraser for the cause-effect relation, the numerical results of resolution enhancement by the post-measurement may intrigue the same philosophical issue of the cause-effect relation, because the post-measurement cannot intervene in the superposition process of the N waves retrospectively. As already known by SNL, however, the statistical ensemble contributes to the resolution enhancement by a factor $\sqrt{K}$. Besides, the Kth power to the N-slit interference fringes was numerically demonstrated to overcome the superresolution in quantum sensing based on N00N states by the same factor $\sqrt{K}$. Thus, the proposed Kth power to the interference fringes of an N-slit interferometer showed the most enhanced resolution in coherence spectroscopy as well as in quantum metrology.

**Methods**

For the Kth-order intensity product in Fig. 1, one output port of the interferometer is divided into K equal ports. This technique of the Kth power to the interference fringe is adapted from the projection measurement in quantum sensing [18-22]. With no phase-dependent projection measurement, all split lights satisfy the statistical ensemble of measurements by Born's rule stating that the intensity is the absolute square of the amplitude. Thus, the K-divided fields are identical in intensity. The Kth power to the measured optical signal is equivalent to the intensity product between the split fields. For the intensity product beyond N=2 such as in an N-slit interferometer or Fabry-Perot interferometer, a photodetector array may be used, otherwise, a single photodetector is scanned.

**Funding:** This research was supported by the MSIT (Ministry of Science and ICT), Korea, under the ITRC (Information Technology Research Center) support program (IITP 2024-2021-0-01810) supervised by the IITP (Institute for Information & Communications Technology Planning & Evaluation). This work was also supported by GIST via the GIST research program in 2024.

**Author contribution:** BSH solely wrote the paper.

**Competing Interests:** The author declares no competing interest.

**Data Availability**

All data generated or analyzed during this study are included in this published article.